# Multistep reversible excitation transfer in a multicomponent rigid solution: I. Calculation of steady-state and time-resolved fluorescence intensities

**Józef Kuśba**[1]

*Independent researcher. Retired at Faculty of Applied Physics and Mathematics, Gdańsk University of Technology, Gdańsk, Poland*

### Abstract

Previously obtained expressions describing the intensity of stationary fluorescence emitted by a multicomponent solution were significantly improved by using matrix calculus. Then, using a similar technique, new expressions describing the decay of the fluorescence intensity of the multicomponent system after pulsed excitation were found. In both of these cases, the effects of the internal filter, the effects of multistep radiative transfer of excitation energy, the possibility of radiative back-transfer, as well as the possibility of changes in the quantum yield of individual components due to radiationless transfer of excitation energy were taken into account. The cases of one-, two- and three-component systems were discussed in detail.

### Abbreviations

| | | | |
|---|---|---|---|
| MCS | multicomponent solution | PDF | probability density function |
| MEE | molecular electronic excitation | RET | radiative excitation transfer |
| NET | nonradiative excitation transfer | SPDF | subnormalized probability density function |

### Symbols and notation

Latin letters

| | |
|---|---|
| $c_i$ | concentration of molecules of the *i*th component. Eqs. (23), (24) |
| $C(\lambda_{\rm ex},\lambda_{\rm em})$ | factor to account for the effect of light absorption in the sample. Eqs. (22) |
| $E^{(o)}(\lambda_{\rm ex},t)$ | vector of effective quantum yield densities of fluorescence of the order $(o)$ Eqs. (45), (51), (55) |
| $F(\lambda_{\rm em})$ | vector of emission spectrum values. Eq. (14) |
| $g$ | optical geometry factor. Eq. (21) |
| $I_n$ | identity matrix of dimension $n\times n$. Eq. (37) |
| $I_\delta(\lambda_{\rm ex},\lambda_{\rm em},t)$ | fluorescence intensity produced by $\delta$-pulse excitation. Eqs. (48), (56) |
| $I_{\rm ex}$ | photon flux density of the continuous excitation beam. Eqs. (21), (39) |
| $I_{\rm ss}(\lambda_{\rm ex},\lambda_{\rm em})$ | fluorescence intensity produced by continuous excitation. Eq. (21) |
| $J_{\rm ex}$ | photon density of the excitation pulse. Eq. (48) |
| $k_i(\lambda)$ | absorption coefficient of the *i*th component. Eq. (10) |
| $K(\lambda_{\rm ex},\lambda_{\rm em},t)$ | time-dependent one-step radiative transfer matrix. Eq. (65) |
| $l$ | thickness of the cuvette. Eqs. (23), (24) |
| $M(\lambda_{\rm ex},\lambda_{\rm em},\lambda)$ | function needed to calculate the matrix $\kappa(\lambda_{\rm ex},\lambda_{\rm em})$. Eqs. (31), (35) |
| $n$ | number of components in the considered fluorescent solution. Sect. 2 |
| $n_r$ | refractive index of the medium. Eq. (22) |
| $N_i^{\rm abs}$ | number of photons absorbed by the *i*th component. Eq. (1) |
| $N_i^{\rm em}$ | number of photons emitted by the *i*th component. Eq. (1) |
| $N_{ij}^{\rm em}$ | number of photons emitted by the *j*th component due to excitation of the *i*th component. Eq. (17) |
| $N_i^{\rm ex}$ | number of excited molecules of the *i*th component produced by the excitation beam. Eq. (3) |
| $R(\lambda_{\rm ex},\lambda_{\rm em})$ | one-step strictly radiative transfer matrix. Eq. (34) |

[1] ORCID: 0000-0002-9697-0751. Electronic mail: jozkusba@pg.edu.pl

$R_{\mathrm{ex}}$ cross-sectional radius of the excitation beam. Sect. 3

$s$ Laplace variable. Eq. (60)

$t$ time. Eq. (42)

$X(\lambda_{\mathrm{ex}})$ vector of relative absorption coefficients. Eqs. (8), (10)

$X^{\star}(\lambda_{\mathrm{ex}})$ vector of relative excitabilities. Eq. (11)

Greek letters

$\alpha$ Napierian absorbance at $\lambda_{\mathrm{ex}}$ wavelength. Eq. (22)

$\beta$ Napierian absorbance at $\lambda_{\mathrm{em}}$ wavelength. Eq. (22)

$\varepsilon_i(\lambda)$ molar absorption coefficient of the $i$th component. Eqs. (23), (24), (36)

$\eta^{(o)}(\lambda_{\mathrm{ex}})$ vector of effective quantum yields of fluorescence of the order $(o)$. Eqs. (19), (26), (38)

$\kappa(\lambda_{\mathrm{ex}},\lambda_{\mathrm{em}})$ one-step radiative transfer matrix. Eq. (26)

$\kappa_{\mathrm{D}}$ diagonal part of the $\kappa$ matrix. Eq. (87)

$\kappa_{\mathrm{U}}$ upper triangular part of the $\kappa$ matrix. Eq. (85)

$\lambda$ wavelength of light. Sect. 2

$\lambda_{\mathrm{ex}}$ excitation wavelength. Sect. 2

$\lambda_{\mathrm{em}}$ fluorescence observation wavelength. Eq. (14)

$\rho$ reflective loss coefficient. Eq. (22)

$\phi$ matrix of photon emission probabilities. Eq. (16)

$\phi_0$ matrix of absolute quantum yields. Eq. (18)

$\phi_i^{\mathrm{app}}(\lambda)$ apparent absolute quantum yield of the $i$th component. Eq. (2)

$\phi^{\star}(\lambda_{\mathrm{ex}})$ matrix of probabilities of active photon absorption. Eq. (12)

$\Phi(t)$ SPDF matrix of photon emission. Eq. (42)

$\omega(\lambda_{\mathrm{ex}},\lambda_{\mathrm{em}})$ multistep radiative transfer matrix. Eq. (39)

$\Omega(\lambda_{\mathrm{ex}},\lambda_{\mathrm{em}},t)$ time-dependent multistep radiative transfer matrix. Eq. (61)

## 1 Introduction

A multicomponent solution (MCS) typically refers to a mixture or solution that contains more than one distinct component or substance. A fluorescent MCS, which we will also call a fluorescent system, is a solution that contains a solvent and two or more fluorescent solutes (fluorophores or fluorescent molecules) dissolved in it. Depending on the type of solvent and the current physical conditions, the fluorescent molecules have more or less mobility. We classify a given MCS as rigid if the diffusive displacements of the fluorescent molecules during their fluorescence lifetime are so small that they do not affect the observed fluorescence properties of the system. Fluorescent MCSs arouse our interest because we either encounter them as already existing in nature, or they appear in certain chemical processes, or they are intentionally created because of their specific properties.

If a multicomponent mixture already exists, we are often interested in its chemical analysis by determining both the types of individual components and their percentages in the mixture. If the components of the mixture are fluorescent, then important information about its composition can be obtained by studying the fluorescence light of the mixture. Many papers have been devoted to this issue [1-6]. One of the main goals of these considerations is to extract pure emission spectra and concentrations of individual components from the recorded data. A comprehensive review of the experimental and computational methods used here is given in [7,8].

Another important reason for analyzing the fluorescence intensity of multicomponent solutions is to study the phenomena of nonradiative transfer of excitation energy between fluorescent molecules. The investigation procedure here usually involves comparing the fluorescence intensity of a multicomponent solution predicted theoretically with the corresponding intensity observed experimentally. In the case of binary systems, the fluorescence intensity of the excitation energy donor and/or acceptor is studied [9-12]. The occurrence of reversible radiationless excitation energy transfer has also been studied in such systems [13,14]. Binary and ternary solutions of organic dyes are often used as lasing media [15-22]. Compared to single-component solutions, this way, in many cases, a significant improvement in performance and expansion of the spectral range of dye lasers was achieved.

The description of the fluorescence intensity emitted by a system of interacting sets of fluorescent molecules is a difficult and complicated undertaking. This is because the fluorescence of each component individually depends on many parameters, and taking into account the interaction of these components further multiplies their number. The primary effect of the interaction of the components is the radiative and nonradiative transfer of excitation energy between them. As a rule, the fluorescence properties of a multicomponent solution are not expressed by linear combinations of the properties of the individual components, but rather their complex functions. In light of the classification given in [8], the MCSs considered in this work should be classified as complex multifluorometric systems.

Early works on the theoretical description of the fluorescence intensity of solutions refer mainly to single-component solutions and observations made at steady state, which is produced by excitation with light of constant intensity. More advanced studies of this issue also deal with the description of the intensity of fluorescence emitted after excitation with a short pulses of light, the so-called time-resolved fluorescence. In theoretical considerations, it is important here to take into account the effects of the inner filter. The basic works in this area belong to Lommel [23,24], Duseberg [25], Jabłoński [26], Weber [27]. Inner filter correction is also the subject of works [28-31]. A basic expression that takes into account the inner filter effect relating to finding quantum efficiency can be found in Förster's monograph [32]. An analogous expression aimed at finding corrected emission spectra was given by Bączyński and Czajkowski [33] for frontal observation and any possible angles of incidence and observation. A good description of the factors affecting the intensity of steady-state and time-resolved fluorescence can be found in [34]. Among the results of the inner filter effect, the formation of secondary fluorescence and higher order emissions are important. Estimated calculations regarding the influence of secondary effects on the mean lifetime and fluorescence anisotropy are included in the work of Galanin [35], while a deeper analysis of phenomenon of radiative excitation transfer (RET) was the subject of works [36-44]. A comprehensive review of the work on the effect of RET on the fluorescence of single-component systems is also given in [45].

The first attempts to describe the spectral distribution of the fluorescence intensity of solutions containing more than one fluorescing component were made in the early 1930s. We refer here to the expression describing the fluorescence of binary solutions given by K. Weber [27]. Later, expressions aimed at describing the fluorescence spectra of such solutions involving energy transfer between components were obtained in works [46-48]. Particularly noteworthy here is the work of Ketskeméty [49], which addressed the RET issue in detail. The expressions obtained in his work were extended to the case of ternary solutions [50], and these in a subsequent step [51] were combined with the results of the work of Bojarski and Domsta [52] on the effect of nonradiative excitation transfer (NET). In another approach to describe the fluorescence properties of the ternary solution [53], the results of the work of [51] were used in the NET part, and the method described in [54] was used in the RET part. The fluorescence properties of the ternary solution were also the subject of the work [6], where methods for decomposing the fluorescence spectrum of such a solution in the presence of a quencher were analyzed.

A natural extension of the description of the fluorescence properties of binary and ternary solutions is the description of the fluorescence properties of solutions with any number of components. A simple expression for the fluorescence intensity emitted by a mixture of mutually non-interacting *n* components is given in Förster's monograph [32], while a description of steady-state fluorescence intensity in a multicomponent system taking into account the transfer of excitation energy between the components is given in [55]. Many aspects of research related to fluorescence analysis of multicomponent systems are addressed in the works of Warner and co-workers. These works include methods for rapid scanning of spectra [56], methods for analysis of multicomponent fluorescence data [1,57], and strategies for data interpretation analysis [3]. A generalized model predicting the fluorescence spectra of a multicomponent system was also proposed in [58]. A review of work related to fluorescence analysis of complex multifluorophore mixtures is given in [7] and [8].

## 2 Relevant parameters of individual components

The subject of our consideration is the fluorescence properties of a solution of *n* different fluorescent components (fluorophores) dissolved in an optically inactive solvent. We assume that the fluorophores do not react with each other, and that each fluorophore individually exhibits a single-exponential fluorescence decay. In our calculations, we will neglect the presence of polarization effects. This means that the results obtained will be applicable in the presence of strong rotational depolarization and/or under "magic angle" excitation-observation conditions. For theoretical considerations, the components are numbered from 1 to *n*. Excitation energy can be exchanged between components through processes such as NET and/or RET. We assume that both of these energy transfer processes in any pair of solution components can be reversible. In our considerations, for any pair of components *i* and *j* (where $j \neq i$ or $j = i$), we take into account both the forward transfer of molecular electronic excitation (MEE) from component *i* to component *j* and the backward transfer from component *j* to component *i*. In addition, we also take into account the fact that the transfer of MEE between components *i* and *j* can be either single-step or multi-step, often taking place with the participation of the other components of the solution. In the latter case, we consider all possible transfer pathways formed by various combinations of fluorophores mediating the transfer of MEE from component *i* to component *j*. The concentrations of the individual components are $c_i$, their absolute quantum yields are $\phi_{0i}$, their fluorescence lifetimes are $\tau_{0i}$, and their molar absorption coefficients for light with a $\lambda$ wavelength are $\varepsilon_i(\lambda)$.

The fluorescent system described above can be excited either with a beam of light of constant intensity or with short pulses of light ($\delta$-pulses), whose duration is much shorter than the fluorescence duration of each MCS component. In either case, we assume that the excitation light is monochromatic and its wavelength is $\lambda_{\mathrm{ex}}$. We understand the absolute quantum yields $\phi_{0i}$ of individual components as ratios of the number of $\underline{N}_i^{\mathrm{em}}$ quanta emitted by the *i*th component to the number of $\underline{N}_i^{\mathrm{abs}}$ molecules absorbed by that component [59].

$$\phi_{0i} = \frac{\underline{N}_i^{\mathrm{em}}}{\underline{N}_i^{\mathrm{abs}}} \tag{1}$$

The underlining in the symbols $\underline{N}_i^{\mathrm{em}}$ and $\underline{N}_i^{\mathrm{abs}}$ means that these quantities refer to the situation when there are no interactions between the molecules of the *i*th component, as well as interactions of the molecules of the *i*th component with the molecules of other components of the solution. According to Vavilov's law, the quantum yield $\phi_{0i}$ remains independent of the wavelength of the excitation light. However, in practice, it is often found that the quantum yield values measured according to expression (1) depend on $\lambda_{\mathrm{ex}}$ [51,60-62]. Under such conditions, the yield calculated using expression (1) does not meet the conditions for absolute quantum yields. For our purposes, we will call it apparent absolute quantum yield and denote it by $\phi_i^{\mathrm{app}}(\lambda)$

$$\phi_i^{\mathrm{app}}(\lambda) = \frac{\underline{N}_i^{\mathrm{em}}(\lambda)}{\underline{N}_i^{\mathrm{abs}}(\lambda)} \tag{2}$$

It can be assumed that the dependence of $\phi_i^{\mathrm{app}}(\lambda)$ on the wavelength of the excitation light is a result of the fact that at certain wavelength ranges the number of excited molecules formed, $\underline{N}_i^{\mathrm{ex}}$, is smaller than the number of absorbed quanta of excitation light, $\underline{N}_i^{\mathrm{abs}}$. This leads to a modified definition of $\phi_{0i}$

$$\phi_{0i} = \frac{\underline{N}_i^{\mathrm{em}}}{\underline{N}_i^{\mathrm{ex}}} \tag{3}$$

After inserting (3) into (2), we obtain

$$\phi_i^{\mathrm{app}}(\lambda) = \frac{\underline{N}_i^{\mathrm{ex}}(\lambda)}{\underline{N}_i^{\mathrm{abs}}(\lambda)}\phi_{0i} = \phi_i^{\star}(\lambda)\,\phi_{0i} \tag{4}$$

Where the magnitude of $\phi_i^{\star}(\lambda)$ given by the expression

$$\phi_i^{\star}(\lambda) = \frac{\underline{N}_i^{\mathrm{ex}}(\lambda)}{\underline{N}_i^{\mathrm{abs}}(\lambda)} \tag{5}$$

represents the probability that absorption of a quantum of light from the excitation beam through a molecule of component $i$ will result in the formation of an excited molecule of that component. We will assume that the values of $\phi_i^{\mathrm{app}}(\lambda)$ over a sufficiently wide range of wavelengths are known, and that the maximum value of $\phi_i^{\mathrm{app}}(\lambda)$ corresponds to $\phi_i^{\star}(\lambda) = 1$. Hence, based on (4), we can write

$$\phi_{0i} = \max\left(\phi_i^{\mathrm{app}}(\lambda)\right) \tag{6}$$

On the other hand, after inserting (6) into (4), we get

$$\phi_i^{\star}(\lambda) = \frac{\phi_i^{\mathrm{app}}(\lambda)}{\max\left(\phi_i^{\mathrm{app}}(\lambda)\right)} \tag{7}$$

The extraction of two quantities from the apparent absolute quantum yield $\phi_i^{\mathrm{app}}(\lambda)$: the pure absolute quantum yield $\phi_{0i}$ and the excitation yield $\phi_i^{\star}(\lambda)$ is important, because only excited molecules can emit photons, or participate in the NET process.

To describe the fluorescence intensity of MCS, we will use a notation in which the properties pertaining to the individual components are expressed by row vectors of dimension $1 \times n$, or by diagonal matrices of dimension $n \times n$, while the properties pertaining to the transfer of MEE between these components are expressed by elements of full square matrices of dimension $n \times n$. When modeling the process of converting the energy of photons of excitation light into the excitation energy of molecules of individual components, it should be noted that, in general, this process must be treated as a complex process, for which it is allowed that not every photon absorbed by a given MCS component results in the formation of an excited molecule of that component. Thus, the probability that a photon absorbed by an MCS was in fact absorbed by the $i$th component of the MCS can be understood as a component of some $n$-dimensional vector $X(\lambda_{\mathrm{ex}})$ of the form

$$X(\lambda_{\mathrm{ex}}) = \left[X_i(\lambda_{\mathrm{ex}})\right]_{1\times n} \tag{8}$$

The values of the individual components $X_i(\lambda_{\mathrm{ex}})$ of this vector can be expressed by the absorption coefficients $k_i(\lambda_{\mathrm{ex}})$ of the individual components

$$k_i(\lambda_{\mathrm{ex}}) = \ln(10)\,\varepsilon_i(\lambda_{\mathrm{ex}})\,c_i \tag{9}$$

according to equation

$$X_i(\lambda_{\mathrm{ex}}) = \frac{k_i(\lambda_{\mathrm{ex}})}{\sum_{i=1}^{n} k_i(\lambda_{\mathrm{ex}})} \tag{10}$$

The probabilities of the appearance of excited states on the molecules of individual MCS components after the act of absorption of a photon from the excitation beam will be determined by the vector

$$X^{\star}(\lambda_{\mathrm{ex}}) = \left[X_i^{\star}(\lambda_{\mathrm{ex}})\right]_{1\times n} \tag{11}$$

such that the value of the $i$th component of this vector is equal to the probability that the absorption of a photon by MCS from light of wavelength $\lambda_{\mathrm{ex}}$ will result in the formation of an excited state on a molecule belonging to the $i$th component of MCS. Therefore, this vector can be called the vector of relative excitabilities of individual components. The vector $X^{\star}(\lambda_{\mathrm{ex}})$ is related to the vector $X(\lambda_{\mathrm{ex}})$ by equation

$$X^{\star}(\lambda_{\mathrm{ex}}) = X(\lambda_{\mathrm{ex}})\,\phi^{\star}(\lambda_{\mathrm{ex}}) \tag{12}$$

where the matrix $\phi^{\star}(\lambda_{\mathrm{ex}})$ is diagonal

$$\phi^{\star}(\lambda_{\mathrm{ex}}) = \mathrm{diag}\left(\phi_1^{\star}(\lambda_{\mathrm{ex}}), \phi_2^{\star}(\lambda_{\mathrm{ex}}), \ldots, \phi_n^{\star}(\lambda_{\mathrm{ex}})\right) \tag{13}$$

and the values of $\phi_i^{\star}(\lambda_{\mathrm{ex}})$ are defined by expressions (5) and/or (7).

Experimental studies typically measure the fluorescence intensity at a selected wavelength, which we denote by $\lambda_{\mathrm{em}}$. This intensity depends on the values of the emission spectra determined for the individual components at the $\lambda_{\mathrm{em}}$, that is, on the $n$-dimensional vector $F(\lambda_{\mathrm{em}})$ defined as

$$F(\lambda_{\mathrm{em}}) = \left[F_i(\lambda_{\mathrm{em}})\right]_{1\times n} \tag{14}$$

We assume here that the individual emission spectra $F_i(\lambda)$ are normalized to unity

$$\int_0^{\infty} F_i(\lambda)\,d\lambda = 1 \tag{15}$$

In this sense, the emission spectrum $F_i(\lambda)$ can be understood as a probability density function (PDF) having the meaning that the product $F_i(\lambda)\,d\lambda$ represents the probability that the photon emitted by $i$th MCS component has a wavelength in the interval $(\lambda, \lambda + d\lambda)$.

## 3 MCS fluorescence intensity generated by continuous excitation

In this section we find an expression describing the steady-state (ss) intensity $I_{ss}(\lambda_{ex},\lambda_{em})$ of fluorescence reaching the detector and emitted by the MCS when excited by light of constant intensity. To begin with, let us note a very important feature in this context, which is the *n*-by-*n* $\phi$ matrix of the form

$$\phi \equiv \left[ \phi_{ij} \right]_{n\times n} \tag{16}$$

In this matrix, the element $\phi_{ij}$ denotes the probability that MEE produced on the *i*th component molecule will be emitted as a light quantum by the *j*th component molecule. We assume here, the concentrations of the components can be arbitrary, and that probability $\phi_{ij}$ is influenced by the processes of spontaneous emission, internal conversion, and NET. For example, if the number of component *i* molecules excited directly by the excitation beam is equal to $N_i^{ex}$, and then the number $N_{ij}^{em} \le N_i^{ex}$ of these excitations is emitted by component *j* molecules in the form of photons, then

$$\phi_{ij} = \frac{N_{ij}^{em}}{N_i^{ex}} \tag{17}$$

When the concentrations of individual MCS components become very small, the $\phi$ matrix becomes the same as the $\phi_0$ diagonal matrix containing the absolute quantum yields of these components

$$\lim_{\substack{c_i\to 0\\ i=1,\ldots,n}} \phi = \phi_0 = \mathrm{diag}\left(\phi_{01},\phi_{02},\ldots,\phi_{0n}\right) \tag{18}$$

In this work, we will consider that the values of the $\phi_{ij}$ elements of the $\phi$ matrix are known. Expressions to calculate the values of $\phi_{ij}$ for assumed values of parameters characterizing a system with any number of components can be found in few works [63,64]. The most common are such expressions for binary systems [12,65-69]. The application of the Markov chain technique to find the values of $\phi_{ij}$ elements for MCSs containing an arbitrary number of components is the subject of our work [70].

When constructing an expression describing the intensity of fluorescence emitted by MCS, it is necessary to take into account the possibility of RET in the described system. The mechanism of RET is that a certain portion of the primary fluorescence light does not go outside the sample, but is absorbed inside it. This is the well-known phenomenon of reabsorption. Primary fluorescence is defined as the part of the total fluorescence that is emitted without RET intermediation, whereas NET intermediation is admissible here. The reabsorbed primary fluorescence generates new excited states, and these are the source of additional emission called secondary fluorescence. This process can be repeated many times, so that in general the observed fluorescence of $I_{ss}(\lambda_{ex},\lambda_{em})$ is the sum of primary fluorescence of $I_{ss}^{(I)}(\lambda_{ex},\lambda_{em})$, secondary fluorescence of $I_{ss}^{(II)}(\lambda_{ex},\lambda_{em})$, tertiary fluorescence of $I_{ss}^{(III)}(\lambda_{ex},\lambda_{em})$, quaternary fluorescence of $I_{ss}^{(IV)}(\lambda_{ex},\lambda_{em})$, etc. In the case of small concentrations of components of a given MCS, combined with the small geometric size of the test sample, it can be assumed that the contribution of the intensity of secondary emission and higher order emission to the total fluorescence intensity is negligibly small. However, the only way to confirm the validity of this assumption for a given MCS is to compare the theoretically estimated magnitudes of these intensities with each other.

Our calculations of the magnitudes of the fluorescence intensities of the various orders emitted by MCS will begin with a description of the magnitude of the primary fluorescence intensity. Of importance here is the vector of effective quantum yields of primary fluorescence of the form

$$\eta^{(I)}(\lambda_{ex}) = \left[ \eta_i^{(I)}(\lambda_{ex}) \right]_{1\times n} \tag{19}$$

The value of the $\eta_i^{(I)}(\lambda_{ex})$ component of this vector is defined as equal to the probability that the absorption of a photon by the entire system from an excitation beam of wavelength $\lambda_{ex}$ will result in the emission of a photon of primary fluorescence by any of the molecules of the component *i*. As in [51] and [55], we will refer to the quantity $\eta_i^{(I)}(\lambda_{ex})$ as the effective quantum yield of primary fluorescence of the *i*th component. Note that in earlier works this quantity was called the "apparent quantum yield of the *i*th component" [60], or "partial quantum yield of the *i*th component" [71]. From the above definitions of $\eta^{(I)}(\lambda_{ex})$, $X^{\star}(\lambda_{ex})$, and $\phi$, it follows that

$$\eta^{(I)}(\lambda_{ex}) = \left[ \sum_{j=1}^{n} X_j^{\star}(\lambda_{ex})\, \phi_{ji} \right]_{1\times n} = X^{\star}(\lambda_{ex})\, \phi \tag{20}$$

Taking into account previous approaches to the problem [32,49,55] we can write an expression describing the intensity of primary fluorescence $I_{ss}^{(I)}(\lambda_{ex},\lambda_{em})$ reaching the detector from the MCS under continuous excitation conditions

$$I_{ss}^{(I)}(\lambda_{ex},\lambda_{em}) = g\, I_{ex}\, C(\lambda_{ex},\lambda_{em})\, \eta^{(I)}(\lambda_{ex}) \left[ F(\lambda_{em}) \right]^{T} \tag{21}$$

In this expression, *g* is a constant, $I_{ex}$ is the photon flux density (photons/m$^2$/s) in the excitation beam, and $C(\lambda_{ex},\lambda_{em})$ is a factor that takes into account the absorptive properties of the sample and the geometry of the measurement system recording $I_{ss}(\lambda_{ex},\lambda_{em})$. From Eq. (21) we see that $I_{ss}$ is the fluorescence photon flux density per unit wavelength interval (photons/s/m$^3$). The expressions for $C(\lambda_{ex},\lambda_{em})$ corresponding to the most commonly used excitation-observation configurations, that is, for front face observation, rear face observation, and right angle observation, can be found in [32] and [39]. In our discussion, we will focus mainly on the frontal observation, since samples of any absorbance value can be examined in this geometry. If a sample of the MCS under test is placed in a parallel-sided cuvette of thickness *l,* then the expression describing the $C(\lambda_{ex},\lambda_{em})$ factor takes the form [39]

$$C(\lambda_{ex},\lambda_{em}) = \frac{\rho}{n_r^2} \frac{\alpha}{\alpha+\beta} \left( 1 - \exp\left[ -(\alpha+\beta) \right] \right) \tag{22}$$

where $\rho$ is the coefficient describing the reflection loss of the excitation beam on the front face of the cuvette, $n_r$ is the refractive index of the medium, while $\alpha$ and $\beta$ are the Napierian absorbances of the sample for the excitation and observation light, respectively. If we know the absorption spectra and the concentrations of the solution components, then $\alpha$ and $\beta$ can be calculated using the expressions

$$\alpha = \ln(10)\frac{l}{\cos\vartheta}\sum_{i=1}^{n}\varepsilon_i(\lambda_{\text{ex}})\,c_i \tag{23}$$

$$\beta = \ln(10)\,l\sum_{i=1}^{n}\varepsilon_i(\lambda_{\text{em}})\,c_i \tag{24}$$

In expression (23), the presence of the $\cos\vartheta$ factor is due to the approximate consideration of the non-perpendicularity of the incidence of the excitation beam on the cuvette [26,33]. This assumes that the angle between the perpendicular to the front wall of the sample and the excitation beam inside the sample is small and equal to $\vartheta$.

An expression describing the intensity of the secondary fluorescence in the case of two-component solution under steady-state conditions was derived by Ketskeméty [49]. His result can be easily generalized to the case of *n* components if one uses matrix notation consistent with that used in Eq. (21). Then we can write

$$I_{\text{ss}}^{(\text{II})}(\lambda_{\text{ex}},\lambda_{\text{em}}) = g\,I_{\text{ex}}\,C(\lambda_{\text{ex}},\lambda_{\text{em}})\,\eta^{(\text{II})}(\lambda_{\text{ex}},\lambda_{\text{em}})\left[F(\lambda_{\text{em}})\right]^T \tag{25}$$

where $\eta^{(\text{II})}(\lambda_{\text{ex}},\lambda_{\text{em}})$ is the vector of effective quantum yields of secondary fluorescence. Analogous to $\eta_i^{(\text{I})}(\lambda_{\text{ex}})$, the value of the $\eta_i^{(\text{II})}(\lambda_{\text{ex}},\lambda_{\text{em}})$ element is equal to the probability that the absorption of a photon by the entire system from an excitation beam with a wavelength of $\lambda_{\text{ex}}$ will result in the emission of a secondary fluorescence photon by any of the component *i* molecules. According to [49], the vector $\eta^{(\text{II})}(\lambda_{\text{ex}},\lambda_{\text{em}})$ can be written in the form

$$\eta^{(\text{II})}(\lambda_{\text{ex}},\lambda_{\text{em}}) = \eta^{(\text{I})}(\lambda_{\text{ex}})\,\kappa(\lambda_{\text{ex}},\lambda_{\text{em}}) \tag{26}$$

where the elements of the matrix

$$\kappa(\lambda_{\text{ex}},\lambda_{\text{em}}) = \left[\kappa_{ij}(\lambda_{\text{ex}},\lambda_{\text{em}})\right]_{n\times n} \tag{27}$$

represent the probabilities of a secondary emission quantum being emitted by the *j*th MCS component as a result of a one-step radiative transfer of MEE from the *i*th component. The physical meaning $\kappa(\lambda_{\text{ex}},\lambda_{\text{em}})$ matrix elements can be determined by analyzing the interrelationships of selected components of expressions (21) and (25). Details of this analysis can be found in Appendix A. It turns out that a given $\kappa_{ij}(\lambda_{\text{ex}},\lambda_{\text{em}})$ element can be interpreted based on two expressions:

$$\kappa_{ij}(\lambda_{\text{ex}},\lambda_{\text{em}}) = \frac{I_{\text{ss}\,ij}^{(\text{II})}(\lambda_{\text{ex}},\lambda_{\text{em}})}{I_{\text{ss}\,j}^{(\text{I})}(\lambda_{\text{ex}},\lambda_{\text{em}})}\,\frac{\eta_j^{(\text{I})}(\lambda_{\text{ex}})}{\eta_i^{(\text{I})}(\lambda_{\text{ex}})} \tag{28}$$

and

$$\kappa_{ij}(\lambda_{\text{ex}},\lambda_{\text{em}}) = \frac{I_{\text{ss}\,ij}^{(\text{II})}(\lambda_{\text{ex}},\lambda_{\text{em}})}{I_{\text{ss}\,i}^{(\text{I})}(\lambda_{\text{ex}},\lambda_{\text{em}})}\,\frac{F_i(\lambda_{\text{em}})}{F_j(\lambda_{\text{em}})} \tag{29}$$

In both expressions, $I_{\text{ss}\,ij}^{(\text{II})}(\lambda_{\text{ex}},\lambda_{\text{em}})$ is that part of the total secondary emission intensity that is emitted by component *j* as a result of RET from component *i*. The $I_{\text{ss}\,j}^{(\text{I})}(\lambda_{\text{ex}},\lambda_{\text{em}})$ and $I_{\text{ss}\,i}^{(\text{I})}(\lambda_{\text{ex}},\lambda_{\text{em}})$ appearing in the denominators of these expressions denote that portion of the total primary emission intensity that is emitted by component *j* or *i*, respectively. When RET occurs between molecules of the same component $(j = i)$ both (28) and (29) take the same form, consistent with that given in [38]

$$\kappa_{ii}(\lambda_{\text{ex}},\lambda_{\text{em}}) = \frac{I_{\text{ss}\,ii}^{(\text{II})}(\lambda_{\text{ex}},\lambda_{\text{em}})}{I_{\text{ss}\,i}^{(\text{I})}(\lambda_{\text{ex}},\lambda_{\text{em}})} \tag{30}$$

From this we see that the $\kappa_{ij}(\lambda_{\text{ex}},\lambda_{\text{em}})$ coefficients are ratios of selected fractions of the observed intensity of primary and secondary fluorescence emitted by the *i*th and *j*th components, however, taking into account the individual absorption or emission capacities of these components. From the works [36,38,49,55] it follows that the kappa matrix can be calculated using the expression

$$\kappa(\lambda_{\text{ex}},\lambda_{\text{em}}) = \int_0^\infty\left[F(\lambda)\right]^T\eta^{(\text{I})}(\lambda)\,M(\lambda_{\text{ex}},\lambda_{\text{em}},\lambda)\,d\lambda \tag{31}$$

where according to (20)

$$\eta^{(\text{I})}(\lambda) = X^\star(\lambda)\;\phi \tag{32}$$

while the function $M(\lambda_{\text{ex}},\lambda_{\text{em}},\lambda)$ determines the spectral probability distribution of the conversion of primary fluorescence quanta to secondary fluorescence quanta under given excitation and observation conditions. Equations (31) and (32) allow the $\kappa_{ij}(\lambda_{\text{ex}},\lambda_{\text{em}})$ matrix to be represented as

$$\kappa(\lambda_{\text{ex}},\lambda_{\text{em}}) = R(\lambda_{\text{ex}},\lambda_{\text{em}})\,\phi \tag{33}$$

where the matrix $R(\lambda_{\text{ex}},\lambda_{\text{em}})$ is given by the expression

$$R(\lambda_{\text{ex}},\lambda_{\text{em}}) = \int_0^\infty\left[F(\lambda)\right]^T X^\star(\lambda)\,M(\lambda_{\text{ex}},\lambda_{\text{em}},\lambda)\,d\lambda \tag{34}$$

The form of expression (33) reflects the fact that the generation of secondary emission photons is, in general, a composite of two stages. The first of these stages is the strictly radiative transfer of a given MEE from the molecules of the *i*th component to the molecules of any MCS component. The second stage involves intramolecular and nonradiative intermolecular processes leading to the emission of this MEE by the molecules of the *j*th component.

The results of the work [38] allow us to conclude that if the MCS sample is placed in a flat-parallel cuvette of thickness *l* and is excited by a cylindrically shaped light beam of radius $R_{\text{ex}}$ then, in the case of observation of the frontal central part of the excitation area, the function $M(\lambda_{\text{ex}},\lambda_{\text{em}},\lambda)$ is equivalent to the function $M(\alpha,\beta,\gamma,m)$ given by

$$\begin{aligned} M(\alpha,\beta,\gamma,m) = {} & \frac{\alpha+\beta}{1-e^{-(\alpha+\beta)}}\,\frac{\gamma}{2}\int_0^1 e^{-\beta u_0}\int_0^1 e^{-\alpha u} \\ & \times\left[\operatorname{Ei}\left(-\gamma\sqrt{m^2+(u-u_0)^2}\right)-\operatorname{Ei}\left(-\gamma\,|\,u-u_0\,|\right)\right]du\,du_0 \end{aligned} \tag{35}$$

where $\alpha$ and $\beta$ depend on $\lambda_{\text{ex}}$ and $\lambda_{\text{em}}$ through equations (23) and (24), respectively, while $\gamma$ depends on $\lambda$ through equation

$$\gamma(\lambda) = \ln(10)\,l\sum_{i=1}^{n}\varepsilon_i(\lambda)\,c_i \tag{36}$$

In equation (35), the parameter *m* is equal to the ratio of the cross-sectional radius of the excitation beam to the thickness of the sample, $m = R_{ex}/l$. Relevant information on the applicability of the function $M(\alpha,\beta,\gamma,m)$ and how to calculate it can be found in [45]. Although the expression (35) may seem complicated, the calculation of its value is not difficult. The simplest procedure here may be to numerically evaluate the double integral occurring in (35). If in our measurement conditions we have $m \gtrsim 8$, then the values of $M(\alpha,\beta,\gamma,m)$ can be calculated much faster by using expressions obtained by analytical transformations of Eq. (35) [41,45]. The source code of the procedures to calculate $M(\alpha,\beta,\gamma,m)$ according to the latter method, written in FORTRAN and Mathcad, is included in the supplementary materials to this article. An example of the results of calculating the function $M(\alpha,\beta,\gamma,m)$ is shown in Figure 1.

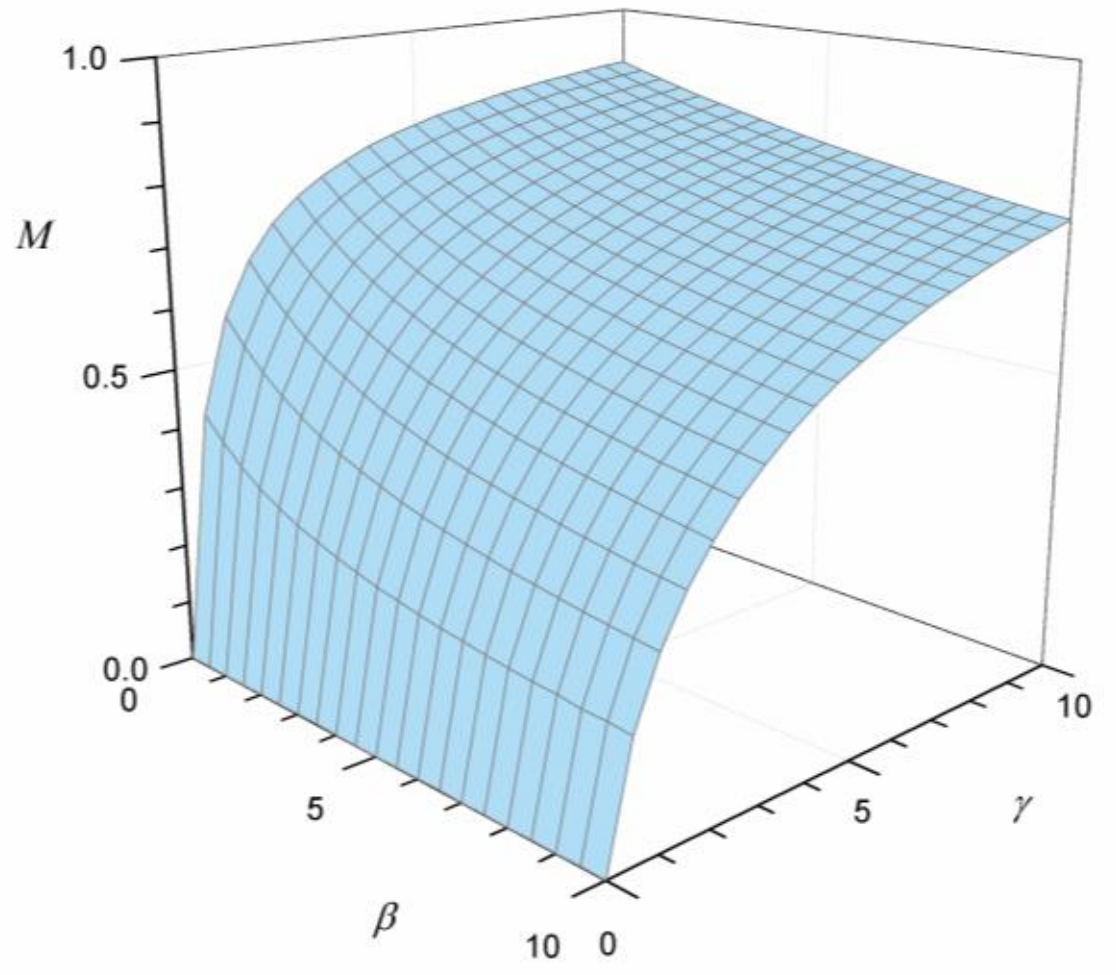

**Figure 1.** The course of the function $M(\alpha,\beta,\gamma,m)$ for $\alpha = 5$ and $m = 10$

As can be seen, the variations of this function throughout the area of applicability are smooth. Additional calculations show that for other real values of the parameters $\alpha$ and $m$, the values of the $M(\alpha,\beta,\gamma,m)$ function change, but the general nature of its course remains the same.

If the intensity of the RET in the considered MCS is not too high, then in the expression describing the fluorescence intensity it is sufficient to consider only primary and secondary emission. Then after summing the expressions (21) and (25), and taking into account (26), we can write

$$I_{ss}^{(I+II)}(\lambda_{ex},\lambda_{em}) = g\, I_{ex}\, C(\lambda_{ex},\lambda_{em})\,\eta^{(I)}(\lambda_{ex}) \times\left[I_n + \kappa(\lambda_{ex},\lambda_{em})\right]\left[F(\lambda_{em})\right]^T \qquad (37)$$

where $I_n$ is an identity matrix of dimension $n \times n$. When the thickness of the sample and/or the concentrations of the fluorescent components it contains are not sufficiently small, a significant contribution of higher order emissions such as tertiary fluorescence, quaternary fluorescence, etc. can be expected in the observed emission. Accurate calculation of the intensity of these higher order emissions is difficult. However, it is relatively easy to make approximate calculations here. For example, for small values of the parameters $\alpha$, $\beta$, and $\gamma$, it can be assumed [38] that the vector of effective quantum yields of fluorescence of the order $(o)$, where $o \in \{\mathrm{III}, \mathrm{IV}, \ldots\}$, is expressed by the vector of effective quantum yields of fluorescence of the order $(o-\mathrm{I})$ according to the recursive relation

$$\eta^{(o)}(\lambda_{ex},\lambda_{em}) = \eta^{(o-1)}(\lambda_{ex},\lambda_{em})\,\kappa(\lambda_{ex},\lambda_{em}) \qquad (38)$$

where $\eta^{(II)}(\lambda_{ex},\lambda_{em})$ is given by Eq. (26), and the $\kappa(\lambda_{ex},\lambda_{em})$ matrix for all orders of emission is the same as for secondary emission. If we express the observed total fluorescence intensity $I_{ss}(\lambda_{ex},\lambda_{em})$ as the sum of the fluorescence intensities of all orders, then after using (38) we can write

$$I_{ss}(\lambda_{ex},\lambda_{em}) = g\, I_{ex}\, C(\lambda_{ex},\lambda_{em}) \times\eta^{(I)}(\lambda_{ex})\,\omega(\lambda_{ex},\lambda_{em})\left[F(\lambda_{em})\right]^T \qquad (39)$$

where the $n \times n$ matrix $\omega(\lambda_{ex},\lambda_{em})$ is a sum of the geometric series generated by consecutive powers of the $\kappa(\lambda_{ex},\lambda_{em})$ matrix

$$\omega(\lambda_{ex},\lambda_{em}) = I_n + \kappa(\lambda_{ex},\lambda_{em}) +\kappa^2(\lambda_{ex},\lambda_{em}) + \kappa^3(\lambda_{ex},\lambda_{em}) + \ldots \qquad (40)$$

It is worth noting that the $\omega(\lambda_{ex},\lambda_{em})$ matrix defined in this way ensures that all possible paths of radiative transfer of excitation energy between the components of the considered MCS can be included in the calculations. In some works [72-74], the series (40) is called the Neumann series. Calculations based on experimental data show that all elements of the matrix $\kappa(\lambda_{ex},\lambda_{obs})$ are nonnegative and less than unity. This makes it possible to suppose that in many experimental cases $\lim_{k\to\infty}\kappa^k = 0$ may occur, which is a condition for the convergence of series (40). Under such conditions, the series (40) can be written in the closed form as [75]

$$\omega(\lambda_{ex},\lambda_{em}) = \left[I_n - \kappa(\lambda_{ex},\lambda_{em})\right]^{-1} \qquad (41)$$

Both expressions (40) and (41) are new in describing the effect of radiative transfer of excitation energy on the fluorescence intensity of a multicomponent system under steady-state conditions. Of particular value here seems to be expression (41) which takes into account the effect of fluorescence of all orders in a simple way. The $\omega(\lambda_{ex},\lambda_{em})$ matrix has not yet been used in describing experimental data on MCSs. However, there are a few papers in the literature that used a description of fluorescence intensity consistent with a limited number of initial terms of the series (40). A deeper analysis of this issue can be found in Appendix B.

## 4 Time-dependent intensity of fluorescence generated by $\delta$-pulse excitation

In the previous section, the use of vector-matrix calculus made it possible to include the contribution of all-order emission in the description of the fluorescence intensity of MCS upon excitation with light of constant intensity. The purpose of this section is, using the same methods, to find an expression describing the time course of the intensity $I_\delta(\lambda_{ex},\lambda_{em},t)$ of the fluorescence emitted by the MCS and observed by the photodetector after $\delta$-pulse excitation. Under pulsed excitation conditions, we assume that at time $t=0$ the MCS under study is illuminated with a $\delta$-pulse of light of wavelength $\lambda_{\mathrm{ex}}$. We will assume that immediately before entering the sample, the surface photon density (photons/m$^2$) in this pulse is $J_{\mathrm{ex}}$. The function $I_\delta(\lambda_{\mathrm{ex}},\lambda_{\mathrm{em}},t)$ determines the temporal distribution of the number of photons emitted toward the detector after the excitation pulse. To achieve our goal, let us first note that a very important feature characterizing the temporal distribution of MCS fluorescence is the matrix function $\Phi(t)$ of the form

$$\Phi(t)=\left[\Phi_{ij}(t)\right]_{n\times n} \tag{42}$$

The elements $\Phi_{ij}(t)$ are functions of time such that the product $\Phi_{ij}(t)\,dt$ is equal to the probability that the excitation of a molecule of component $i$ at $t=0$ will result in the emission of a photon by a molecule of component $j$ in the time interval $(t,t+dt)$. The functions $\Phi_{ij}(t)$ are supported on the interval $[0,\infty)$ and by definition are zero for $t<0$. In determining the value of the function $\Phi_{ij}(t)$ one should take into account the processes of multistep reversible nonradiative energy transfer, including both heterotransfer and homotransfer. We will assume here that the matrix of functions $\Phi(t)$ does not contain information about the effect of RET on the fluorescence of the considered system. For a given MCS, the functions $\Phi_{ij}(t)$ can be determined experimentally only in very simple systems with a minimum number of components and for certain wavelength ranges of $\lambda_{\mathrm{ex}}$ and $\lambda_{\mathrm{em}}$. In general, it can be assumed that the courses of these functions can be determined theoretically, after adopting an appropriate excitation energy transfer mechanism and using an appropriate computational model. Potentially, the resulting expressions can also take into account the presence of material diffusion in the MCS under consideration. Such calculations, can be found, for example, in works [52,64,66-68]. A new approach to calculating the function $\Phi_{ij}(t)$ using the formalism of Markov processes is presented in our next work [70]. In the framework of the present work, we will assume that the matrix of functions $\Phi(t)$ is known. Note that the elements $\Phi_{ij}(t)$ describe probability density distributions but are not normalized to unity. It follows from the above assumptions that their normalization constants are the $\phi_{ij}$ elements of the matrix $\phi$ defined by Eq. (16), that is, we can write

$$\int_0^\infty \Phi_{ij}(t)\,dt=\phi_{ij} \tag{43}$$

or in matrix form

$$\int_0^\infty \Phi(t)\,dt=\phi=\left[\phi_{ij}\right]_{n\times n} \tag{44}$$

We classify $\Phi_{ij}(t)$ functions as subnormalized PDFs (SPDFs), due to the fact that they have all PDF attributes except the condition of normalization to unity. In the rest of this paper, the $\Phi_{ij}(t)$ functions will be referred to as photon emission SPDFs. The unit of $\Phi_{ij}(t)$ is 1/s.

After defining the matrix $\Phi(t)$, to describe MCS fluorescence, we can introduce the vector of effective quantum yield densities of primary fluorescence, $E^{(\mathrm{I})}(\lambda_{\mathrm{ex}},t)$, of the form

$$E^{(\mathrm{I})}(\lambda_{\mathrm{ex}},t)=\left[E_i^{(\mathrm{I})}(\lambda_{\mathrm{ex}},t)\right]_{1\times n} \tag{45}$$

The elements $E_i^{(\mathrm{I})}(\lambda_{\mathrm{ex}},t)$ of this vector are such that the product $E_i^{(\mathrm{I})}(\lambda_{\mathrm{ex}},t)\,dt$ is equal to the probability that a photon absorbed by MCS from an excitation light beam of wavelength $\lambda_{\mathrm{ex}}$ at time $t=0$, will cause a light quantum to be emitted by molecule of component $i$ in the time interval $(t,t+dt)$. From the above definitions of $E^{(\mathrm{I})}(\lambda_{\mathrm{ex}},t)$, $X(\lambda_{\mathrm{ex}})$, and $\Phi(t)$, it follows that

$$E^{(\mathrm{I})}(\lambda_{\mathrm{ex}},t)=\left[\sum_{j=1}^{n} X_j^{\star}(\lambda_{\mathrm{ex}})\,\Phi_{ji}(t)\right]_{1\times n}=X^{\star}(\lambda_{\mathrm{ex}})\,\Phi(t) \tag{46}$$

Given equations (44) and (20), it is easy to see that the integral of the element $E_i^{(\mathrm{I})}(\lambda_{\mathrm{ex}},t)$ taken over time from zero to infinity is equal to $\eta_i^{(\mathrm{I})}(\lambda_{\mathrm{ex}})$

$$\int_0^\infty E_i^{(\mathrm{I})}(\lambda_{\mathrm{ex}},t)\,dt=\eta_i^{(\mathrm{I})}(\lambda_{\mathrm{ex}}) \tag{47}$$

As in the case up to $\Phi_{ij}(t)$, the unit of $E_i^{(\mathrm{I})}(\lambda_{\mathrm{ex}},t)$ is 1/s.

We will begin the construction of the expression for the function $I_\delta(\lambda_{\mathrm{ex}},\lambda_{\mathrm{em}},t)$ by considering the simplest case, which only involves primary fluorescence and is described by the function $I_\delta^{(\mathrm{I})}(\lambda_{\mathrm{ex}},\lambda_{\mathrm{em}},t)$. It can be predicted that the structure of the expression for the function $I_\delta^{(\mathrm{I})}(\lambda_{\mathrm{ex}},\lambda_{\mathrm{em}},t)$ corresponds to the structure of the expression (21), in which the photon flux density of the excitation beam $I_{\mathrm{ex}}$ is replaced by the photon density of the excitation pulse $J_{\mathrm{ex}}$, and the vector $\eta^{(\mathrm{I})}(\lambda_{\mathrm{ex}})$ is replaced by the time-dependent vector $E^{(\mathrm{I})}(\lambda_{\mathrm{ex}},t)$

$$I_\delta^{(\mathrm{I})}(\lambda_{\mathrm{ex}},\lambda_{\mathrm{em}},t)=g\,J_{\mathrm{ex}}\,C(\lambda_{\mathrm{ex}},\lambda_{\mathrm{em}})\,E^{(\mathrm{I})}(\lambda_{\mathrm{ex}},t)\left[F(\lambda_{\mathrm{em}})\right]^T \tag{48}$$

Since the unit of $J_{\mathrm{ex}}$ is photon/m$^2$, so we see here that, as with $I_{\mathrm{ss}}^{(\mathrm{I})}(\lambda_{\mathrm{ex}},\lambda_{\mathrm{em}})$, the $I_\delta^{(\mathrm{I})}(\lambda_{\mathrm{ex}},\lambda_{\mathrm{em}},t)$ fluorescence intensity is expressed in photons/s/m$^3$. Note that the expressions (48) and (21) satisfy the relation

$$\int_0^\infty I_\delta(\lambda_{\mathrm{ex}},\lambda_{\mathrm{em}},t)\,dt=\frac{J_{\mathrm{ex}}}{I_{\mathrm{ex}}}\,I_{\mathrm{ss}}(\lambda_{\mathrm{ex}},\lambda_{\mathrm{em}}) \tag{49}$$

found in Appendix C. On the same principle, we predict that the expression describing the time course of secondary fluorescence is given by expression

$$\begin{aligned} I_\delta^{(\mathrm{II})}(\lambda_{\mathrm{ex}},\lambda_{\mathrm{em}},t)&=g\,J_{\mathrm{ex}}\,C(\lambda_{\mathrm{ex}},\lambda_{\mathrm{em}})\\ &\times E^{(\mathrm{II})}(\lambda_{\mathrm{ex}},\lambda_{\mathrm{em}},t)\left[F(\lambda_{\mathrm{em}})\right]^T \end{aligned} \tag{50}$$

where the vector $E^{(\mathrm{II})}(\lambda_{\mathrm{ex}},\lambda_{\mathrm{em}},t)$, which we call the vector of effective quantum yield densities of secondary fluorescence is defined by the equation analogous to (26)

$$E^{(\mathrm{II})}(\lambda_{\mathrm{ex}},\lambda_{\mathrm{em}},t)=\left(E^{(\mathrm{I})}(\lambda_{\mathrm{ex}},\cdot)*K(\lambda_{\mathrm{ex}},\lambda_{\mathrm{em}},\cdot)\right)(t) \tag{51}$$

It is worth noting that after replacing $\eta^{(\mathrm{I})}(\lambda_{\mathrm{ex}})$ by $E^{(\mathrm{I})}(\lambda_{\mathrm{ex}},t)$ and $\kappa(\lambda_{\mathrm{ex}},\lambda_{\mathrm{em}})$ by $K(\lambda_{\mathrm{ex}},\lambda_{\mathrm{em}},t)$, the resulting product of time-dependent functions is treated as convolution of these functions. Such a procedure ensures that the time integral of equation (51) taken from zero to infinity gives equation (26). The matrix $K(\lambda_{\mathrm{ex}},\lambda_{\mathrm{em}},t)$ describes the temporal effect of radiative transfer of MEE on the course of MCS fluorescence decay. To calculate the elements of this matrix, we will assume that the MCS sample is small enough and the fluorescence lifetime of the individual MCS components is large enough that the effects associated with the fluorescence transit time between molecules in the RET process can be neglected. Then the kappa matrix can be expressed analogously to the $\kappa(\lambda_{\mathrm{ex}},\lambda_{\mathrm{em}})$ matrix in Eq. (31)

$$K(\lambda_{\mathrm{ex}},\lambda_{\mathrm{em}},t)=\int\limits_0^\infty \left[F(\lambda)\right]^T E^{(\mathrm{I})}(\lambda,t)\,M(\lambda_{\mathrm{ex}},\lambda_{\mathrm{em}},\lambda)\,d\lambda \qquad (52)$$

Taking into account Eq. (46), it is easy to see that equation (52) can also be written in the form of

$$K(\lambda_{\mathrm{ex}},\lambda_{\mathrm{em}},t)=R(\lambda_{\mathrm{ex}},\lambda_{\mathrm{em}})\,\Phi(t) \qquad (53)$$

where the matrix $R(\lambda_{\mathrm{ex}},\lambda_{\mathrm{em}})$ is given by Eq. (34). From Eqs. (52), (47), and (31) also follows the relation

$$\int\limits_0^\infty K(\lambda_{\mathrm{ex}},\lambda_{\mathrm{em}},t)\,dt=\kappa(\lambda_{\mathrm{ex}},\lambda_{\mathrm{em}}) \qquad (54)$$

To determine the fluorescence intensities of the higher orders, that is, when $o\geq \mathrm{III}$, we will use a recursive approximation for the vector of effective quantum yield densities of fluorescence of order $(o)$ analogous to the vector described by equation (38)

$$E^{(o)}(\lambda_{\mathrm{ex}},\lambda_{\mathrm{em}},t)=\left(E^{(o-\mathrm{I})}(\lambda_{\mathrm{ex}},\lambda_{\mathrm{em}},\cdot)*K(\lambda_{\mathrm{ex}},\lambda_{\mathrm{em}},\cdot)\right)(t) \qquad (55)$$

where $E^{(\mathrm{II})}(\lambda_{\mathrm{ex}},\lambda_{\mathrm{em}},t)$ is given by Eq. (51).The observed fluorescence intensity course is the sum of the intensity courses of the individual orders. As a result of this summation, we obtain

$$\begin{aligned} I_\delta(\lambda_{\mathrm{ex}},\lambda_{\mathrm{em}},t)&=g\,J_{\mathrm{ex}}\,C(\lambda_{\mathrm{ex}},\lambda_{\mathrm{em}})\\ &\times\left(E^{(\mathrm{I})}(\lambda_{\mathrm{ex}},\cdot)*\Omega(\lambda_{\mathrm{ex}},\lambda_{\mathrm{em}},\cdot)\right)(t)\left[F(\lambda_{\mathrm{em}})\right]^T \end{aligned} \qquad (56)$$

where

$$\begin{aligned} \Omega(\lambda_{\mathrm{ex}},\lambda_{\mathrm{em}},t)&=I_n\,\delta(t)+K(\lambda_{\mathrm{ex}},\lambda_{\mathrm{em}},t)\\ &+K^{*2}(\lambda_{\mathrm{ex}},\lambda_{\mathrm{em}},t)\;+K^{*3}(\lambda_{\mathrm{ex}},\lambda_{\mathrm{em}},t)\;+\ldots \end{aligned} \qquad (57)$$

In Eq. (57), $K^{*k}(\lambda_{\mathrm{ex}},\lambda_{\mathrm{em}},t)$ is the *k*th convolution power of the matrix $K(\lambda_{\mathrm{ex}},\lambda_{\mathrm{em}},t)$.

$$\begin{aligned} &K^{*k}(\lambda_{\mathrm{ex}},\lambda_{\mathrm{em}},t)\\ &\equiv\underbrace{\left(K(\lambda_{\mathrm{ex}},\lambda_{\mathrm{em}},\cdot)*K(\lambda_{\mathrm{ex}},\lambda_{\mathrm{em}},\cdot)*\ldots*K(\lambda_{\mathrm{ex}},\lambda_{\mathrm{em}},\cdot)\right)}_{k\ \text{members}}(t) \end{aligned} \qquad (58)$$

Based on (57), (54), and (40), we also have

$$\int\limits_0^\infty \Omega(\lambda_{\mathrm{ex}},\lambda_{\mathrm{em}},t)\,dt=\omega(\lambda_{\mathrm{ex}},\lambda_{\mathrm{em}}) \qquad (59)$$

The procedure for calculating the function $I_\delta(\lambda_{\mathrm{ex}},\lambda_{\mathrm{em}},t)$, illustrated by equations (56), (57), and (51), is greatly simplified if, instead of a time-dependent function, one first calculates its Laplace transform $\hat{I}_\delta(\lambda_{\mathrm{ex}},\lambda_{\mathrm{em}},s)\equiv\mathfrak{L}\left[I_\delta(\lambda_{\mathrm{ex}},\lambda_{\mathrm{em}},t)\right]$, where

$$\mathfrak{L}\left(f(t)\right)=\hat{f}(s)=\int_0^\infty \exp(-st)\,f(t)\,dt \qquad (60)$$

In Laplace space, equation (56) takes the form

$$\begin{aligned} \hat{I}_\delta(\lambda_{\mathrm{ex}},\lambda_{\mathrm{em}},s)&=g\,J_{\mathrm{ex}}\,C(\lambda_{\mathrm{ex}},\lambda_{\mathrm{em}})\\ &\times\hat{E}^{(\mathrm{I})}(\lambda_{\mathrm{ex}},s)\,\hat{\Omega}(\lambda_{\mathrm{ex}},\lambda_{\mathrm{em}},s)\left[F(\lambda_{\mathrm{em}})\right]^T \end{aligned} \qquad (61)$$

where $\hat{E}^{(\mathrm{I})}(\lambda_{\mathrm{ex}},s)$ is the Laplace transform of the vector $E^{(\mathrm{I})}(\lambda_{\mathrm{ex}},t)$ defined by Eq. (46)

$$\hat{E}^{(\mathrm{I})}(\lambda_{\mathrm{ex}},s)=X^{\star}(\lambda_{\mathrm{ex}})\;\hat{\Phi}(s) \qquad (62)$$

From Eqs. (47) and (60), we also see that

$$\hat{E}^{(\mathrm{I})}(\lambda_{\mathrm{ex}},s=0)=\eta^{(\mathrm{I})}(\lambda_{\mathrm{ex}}) \qquad (63)$$

In Eq. (61), $\hat{\Omega}(\lambda_{\mathrm{ex}},\lambda_{\mathrm{em}},s)$ is the Laplace transform of the matrix $\Omega(\lambda_{\mathrm{ex}},\lambda_{\mathrm{em}},t)$ defined by Eq. (57). After the Laplace transformation, the convolution powers reduce to ordinary powers, so that we can write

$$\begin{aligned} \hat{\Omega}(\lambda_{\mathrm{ex}},\lambda_{\mathrm{em}},s)&=I_n+\hat{K}(\lambda_{\mathrm{ex}},\lambda_{\mathrm{em}},s)\\ &+\hat{K}^2(\lambda_{\mathrm{ex}},\lambda_{\mathrm{em}},s)+\hat{K}^3(\lambda_{\mathrm{ex}},\lambda_{\mathrm{em}},s)+\ldots \end{aligned} \qquad (64)$$

where

$$\hat{K}(\lambda_{\mathrm{ex}},\lambda_{\mathrm{em}},s)=\int\limits_0^\infty F(\lambda)\,\hat{E}^{(\mathrm{I})}(\lambda,s)\,M(\lambda_{\mathrm{ex}},\lambda_{\mathrm{em}},\lambda)\,d\lambda \qquad (65)$$

From Eqs. (65), (63), and (31), it follows that there is a relation

$$\hat{K}(\lambda_{\mathrm{ex}},\lambda_{\mathrm{em}},s=0)=\kappa(\lambda_{\mathrm{ex}},\lambda_{\mathrm{em}}) \qquad (66)$$

Taking into account the fact that functions $\hat{K}(\lambda_{\mathrm{ex}},\lambda_{\mathrm{em}},s)$ decrease with increasing values of the variable *s*, and considering the discussion of the values of the elements of the $\kappa(\lambda_{\mathrm{ex}},\lambda_{\mathrm{em}})$ matrix given after equation (40), we conclude that under typical experimental conditions, for any value of *s*, the series (64) converges. Then, as in the case of series (40), we can write

$$\hat{\Omega}(\lambda_{\mathrm{ex}},\lambda_{\mathrm{em}},s)=\left[I_n-\hat{K}(\lambda_{\mathrm{ex}},\lambda_{\mathrm{em}},s)\right]^{-1} \qquad (67)$$

Note that due to the nature of the fluorescence phenomenon, the functions $\Phi_{ij}(t)$ contained in the matrix $\Phi(t)$ and in the vector $E^{(\mathrm{I})}(\lambda_{\mathrm{ex}},t)$, functions $K_{ij}(\lambda_{\mathrm{ex}},\lambda_{\mathrm{em}},t)$, as well as the entire function $I_\delta(\lambda_{\mathrm{ex}},\lambda_{\mathrm{em}},t)$ must be bounded, nonnegative, and should decrease to zero when *t* goes to infinity. Thus, one can assume that Laplace transforms of these functions exist.

By calculating $\hat{\Omega}(\lambda_{\mathrm{ex}},\lambda_{\mathrm{em}},s)$ from Eq. (67) and inserting the resulting values into equation (61), we find the values of the Laplace transform of the intensity of $\hat{I}_\delta(\lambda_{\mathrm{ex}},\lambda_{\mathrm{em}},s)$ taking into account the primary emission and emissions of all higher orders. These values can then be inverted to time space using any of the numerical methods [76]. Expressions (61)-(67) are fundamental to the theoretical calculation of the time course of the fluorescence intensity of MCS. All parameters appearing on its right-hand side can be determined either directly experimentally or after some additional theoretical considerations. In the particular case of a homogeneous system, consisting of just one component, equation (61) after taking into account (67), (65), (46), (31), and (20) reduces to

$$\hat{I}_\delta(\lambda_{ex},\lambda_{em},s) = g\,J_{ex}\,C(\lambda_{ex},\lambda_{em}) \times\phi_1^\star(\lambda_{ex})\frac{\hat{\Phi}_{11}(s)}{1-\kappa_{11}(\lambda_{ex},\lambda_{em})\hat{\Phi}_{11}(s)/\phi_{11}}F_1(\lambda_{em}) \tag{68}$$

which, as shown in Appendix D, is consistent with the previously obtained equation (30) in [45].

Relationships (44) and (59) can also be written as

$$\phi = \hat{\Phi}(s=0) \tag{69}$$

$$\omega(\lambda_{ex},\lambda_{em}) = \hat{\Omega}(\lambda_{ex},\lambda_{em},s=0) \tag{70}$$

The latter equations, together with equation (63), become useful for calculating steady-state fluorescence parameters when the Laplace transforms of the time characteristics of the fluorescence emitted after $\delta$-pulse excitation are known.

## 5 Summary

The most important achievements of this work are equations (39) and (61). Equation (39) describes the MCS fluorescence intensity under excitation by a beam of light of constant intensity, and equation (61) describes the MCS fluorescence intensity under excitation by delta pulses. In both cases, the possibility of both radiative and non-radiative excitation energy transfer in the described system is taken into account. Almost all the data needed for the calculations come from direct measurements. The exceptions to this are the quantities that depend on the radiationless transfer and which are the elements of the $\phi$ and $\hat{\Phi}(s)$ matrices. To obtain them, additional calculations must be performed, e.g., such as those described in [70]. The application of the matrix formalism to the description of RET in MCS has made it possible to obtain expressions that more completely than before describe the effect of higher-order fluorescence on the observed total fluorescence intensity of the system.

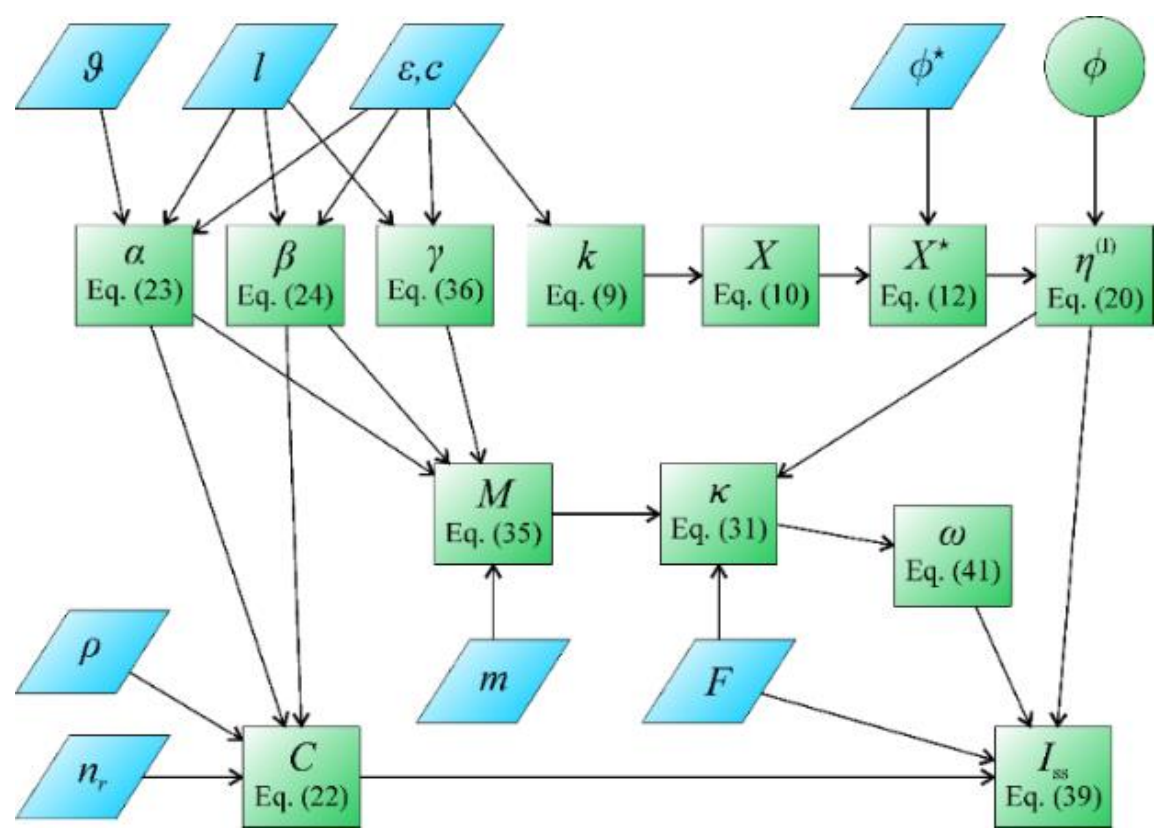

**Figure. 2.** Block diagram of the course of calculation of $I_{ss}(\lambda_{ex},\lambda_{em})$ values according to expression (39). The blue slanted quadrangles represent experimentally determined data, the green circle – the values of the elements of the $\phi$ matrix possible to calculate by the methods discussed in the paper [70]. Green rectangles illustrate the expressions provided in this paper.

The expression (39) is a supplemented and improved version of the equation given earlier [55]. A block diagram of all the calculations that need to be performed before finally using equation (39) is shown in Fig. 2. The calculations illustrated by the green block located in the lower left corner of the diagram ($C$, Eq. (22)) refer to the internal filter effect, the calculations illustrated by the blocks located in the upper right part of the diagram refer to primary fluorescence, and the calculations contained in the blocks located on the diagonal of the diagram refer to secondary fluorescence.

The expression (61) is new. It shows for the first time what is the simultaneous effect of RET and NET on the observed time courses of MCS fluorescence intensity after pulsed excitation. A block diagram of the calculations that need to be performed to use this equation is shown in Fig. 3. These calculations are very similar to those needed to calculate $I_{ss}$. In particular, the values of the function $M(\alpha,\beta,\gamma,m)$ in both cases are calculated in the same way. The values of the functions calculated in the orange blocks are the values of the corresponding Laplace transforms, but the difficulty of these calculations is no greater than in the analogous blocks shown in Fig. 2.

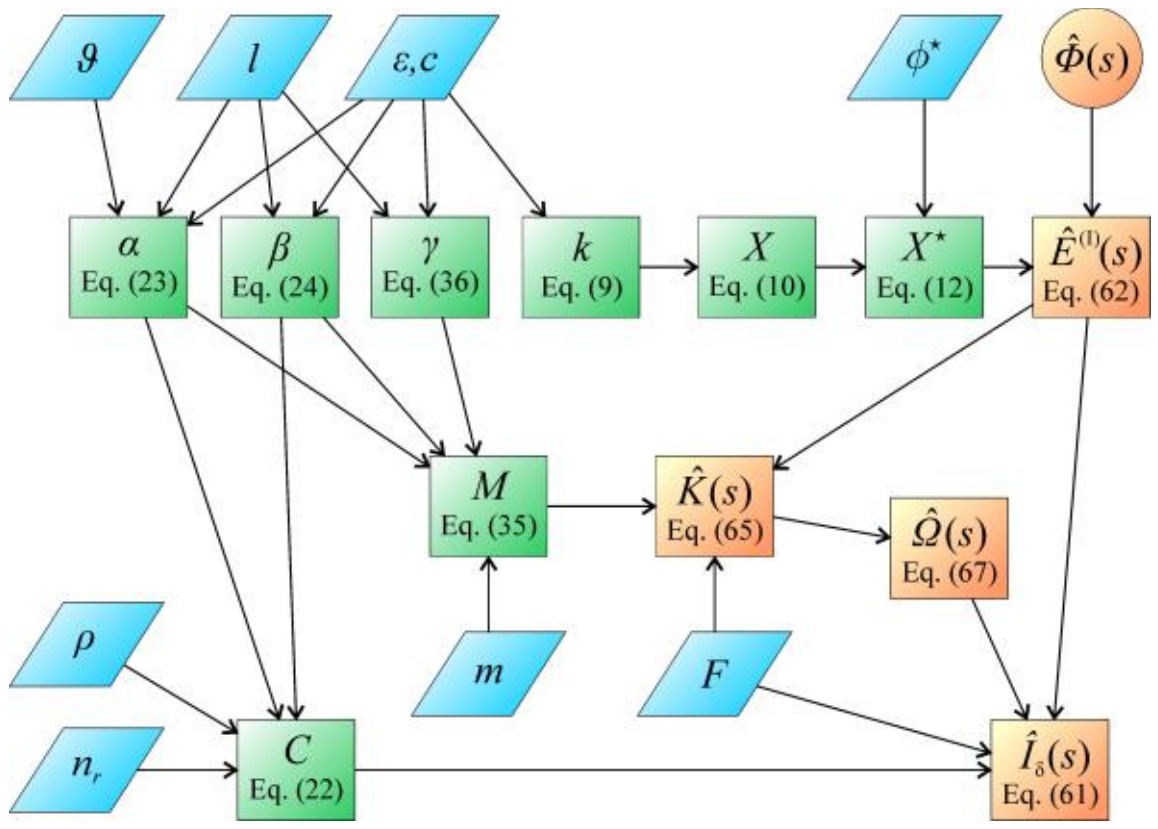

**Figure 3**. Block diagram of the course of $\hat{I}_\delta(\lambda_{ex},\lambda_{em},s)$ calculations according to expression (61). The blue slanted quadrangles represent experimentally determined data. The orange circle contains the values of the $\hat{\Phi}_{ij}(s)$ function that can be calculated by the methods discussed in the paper [70]. The green rectangles illustrate the expressions provided in this paper, which are the same as those used to calculate $I_{ss}(\lambda_{ex},\lambda_{em})$. Orange rectangles indicate expressions that relate to the calculation of Laplace transforms of the time courses of the quantities $E^{(1)}(t)$, $K(t)$, $\Omega(t)$, and $I_\delta(t)$.

In order to find $I_\delta(t)$ values from the calculated $\hat{I}_\delta(s)$ values, one can use any of the numerical methods for inverting Laplace transforms. An exhaustive overview of these methods is given, for example, in [76]. From our preliminary calculations, it appears that the method developed by Stehfest [77,78] may be relatively easy and sufficiently accurate here.

## Appendix A: Physical meaning of the quantity $\kappa_{ij}$

Expression (21) can be rewritten as

$$I_{ss}^{(1)}(\lambda_{ex},\lambda_{em}) = \sum_{k=1}^{n} I_{ss\,k}^{(1)}(\lambda_{ex},\lambda_{em}) \tag{71}$$

where $I_{\mathrm{ss}k}^{(\mathrm{I})}(\lambda_{\mathrm{ex}},\lambda_{\mathrm{em}})$ is that part of the primary emission intensity of the system that is emitted by the molecules of the *k*th component

$$I_{\mathrm{ss}k}^{(\mathrm{I})}(\lambda_{\mathrm{ex}},\lambda_{\mathrm{em}}) = W(\lambda_{\mathrm{ex}},\lambda_{\mathrm{em}})\,\eta_k^{(\mathrm{I})}(\lambda_{\mathrm{ex}})\,F_k(\lambda_{\mathrm{em}}) \tag{72}$$

and $W(\lambda_{\mathrm{ex}},\lambda_{\mathrm{em}}) = g\,I_{\mathrm{ex}}\,C(\lambda_{\mathrm{ex}},\lambda_{\mathrm{em}})$. Similarly, expression (25) can be rewritten as

$$I_{\mathrm{ss}}^{(\mathrm{II})}(\lambda_{\mathrm{ex}},\lambda_{\mathrm{em}}) = \sum_{i=1}^{n}\sum_{j=1}^{n} I_{\mathrm{ss}ij}^{(\mathrm{II})}(\lambda_{\mathrm{ex}},\lambda_{\mathrm{em}}) \tag{73}$$

where $I_{\mathrm{ss}ij}^{(\mathrm{II})}(\lambda_{\mathrm{ex}},\lambda_{\mathrm{em}})$ is that part of the secondary emission of the system which is emitted by molecules of the *j*th component due to RET from molecules of the *i*th component

$$I_{\mathrm{ss}ij}^{(\mathrm{II})}(\lambda_{\mathrm{ex}},\lambda_{\mathrm{em}}) = W(\lambda_{\mathrm{ex}},\lambda_{\mathrm{em}})\,\eta_i^{(\mathrm{I})}(\lambda_{\mathrm{ex}})\,\kappa_{ij}(\lambda_{\mathrm{ex}},\lambda_{\mathrm{em}})\,F_j(\lambda_{\mathrm{em}}) \tag{74}$$

If $k = j$, then it follows from expressions (72) and (74) that

$$\kappa_{ij}(\lambda_{\mathrm{ex}},\lambda_{\mathrm{em}}) = \frac{I_{\mathrm{ss}ij}^{(\mathrm{II})}(\lambda_{\mathrm{ex}},\lambda_{\mathrm{em}})}{I_{\mathrm{ss}\,j}^{(\mathrm{I})}(\lambda_{\mathrm{ex}},\lambda_{\mathrm{em}})}\,\frac{\eta_j^{(\mathrm{I})}(\lambda_{\mathrm{ex}})}{\eta_i^{(\mathrm{I})}(\lambda_{\mathrm{ex}})} \tag{75}$$

If $k = i$, then it follows from expressions (72) and (74) that

$$\kappa_{ij}(\lambda_{\mathrm{ex}},\lambda_{\mathrm{em}}) = \frac{I_{\mathrm{ss}ij}^{(\mathrm{II})}(\lambda_{\mathrm{ex}},\lambda_{\mathrm{em}})}{I_{\mathrm{ss}i}^{(\mathrm{I})}(\lambda_{\mathrm{ex}},\lambda_{\mathrm{em}})}\,\frac{F_i(\lambda_{\mathrm{em}})}{F_j(\lambda_{\mathrm{em}})} \tag{76}$$

## Appendix B: Forms of the $\omega$ matrix for the simplest systems

### *One-component system*

For a one-component system, the $\kappa$ matrix contains only one element $\kappa \equiv \kappa_{11}$, which means that the $\omega$ matrix also contains only one element of the form

$$\omega = \frac{1}{1-\kappa} \tag{77}$$

Here we have a full agreement of equation (77) with expression (2) obtained for the same case in the work of Budó and Ketskeméty [37].

### *Two-component system*

For a binary system, the $\kappa$ matrix contains four elements

$$\kappa = \begin{bmatrix} \kappa_{11} & \kappa_{12} \\ \kappa_{21} & \kappa_{22} \end{bmatrix} \tag{78}$$

and the $\omega$ matrix calculated from Eq. (41) takes the form

$$\omega = \frac{1}{(1-\kappa_{11})(1-\kappa_{22})-\kappa_{12}\kappa_{21}} \begin{bmatrix} 1-\kappa_{22} & \kappa_{12} \\ \kappa_{21} & 1-\kappa_{11} \end{bmatrix} \tag{79}$$

From the paper [49] devoted to the same issue, we conclude that instead of the $\omega$ matrix there was used the $\omega'$ matrix of the form

$$\omega' = \begin{bmatrix} 1+\kappa_{11} & \kappa_{12} \\ \kappa_{21} & 1+\kappa_{22} \end{bmatrix} \tag{80}$$

It is easy to see that the reason for the inconsistency of expressions (80) and (79) is that only the first two components of the series (40) were considered in determining the $\omega'$ matrix

$$\omega' = \omega^{(\mathrm{I+II})} = I_2 + \kappa \tag{81}$$

Matrix (81) was also used to describe the fluorescence intensity of the binary solution in paper [71].

### *Ternary system*

In the case of ternary system, the $\kappa$ matrix contains nine elements

$$\kappa = \begin{bmatrix} \kappa_{11} & \kappa_{12} & \kappa_{13} \\ \kappa_{21} & \kappa_{22} & \kappa_{23} \\ \kappa_{31} & \kappa_{32} & \kappa_{33} \end{bmatrix} \tag{82}$$

and then, according to expression (41), the $\omega$ matrix takes a form $\omega = \left[\omega_{ij}\right]_{3\times 3}$, where

$$\left.\begin{aligned} \omega_{ii} &= \frac{1}{d}\left[(1-\kappa_{jj})(1-\kappa_{kk}) - \kappa_{jk}\kappa_{kj}\right] \\ \omega_{ij} &= \frac{1}{d}\left[\kappa_{ij}(1-\kappa_{kk}) + \kappa_{ik}\kappa_{kj}\right] \end{aligned}\right\} \quad \begin{aligned} &i,j,k = 1,2,3 \\ &k \neq j \neq i \end{aligned} \tag{83}$$

and

$$\begin{aligned} d &= (1-\kappa_{11})(1-\kappa_{22})(1-\kappa_{33}) \\ &\quad -(1-\kappa_{11})\kappa_{23}\kappa_{32} - (1-\kappa_{22})\kappa_{13}\kappa_{31} - (1-\kappa_{33})\kappa_{12}\kappa_{21} \\ &\quad -\kappa_{12}\kappa_{23}\kappa_{31} - \kappa_{13}\kappa_{32}\kappa_{21} \end{aligned} \tag{84}$$

The matrix (83) is new and therefore has not yet been used when describing experimental data on the fluorescence intensity of a ternary solution.

In papers [50] and [51], the fluorescence of specific ternary systems was studied, in which energy transfer from component *i* to component *j* was not possible if $i > j$. Under such conditions, the kappa matrix takes the form of upper triangular matrix $\kappa_{\mathrm{U}}$

$$\kappa_{\mathrm{U}} = \begin{bmatrix} \kappa_{11} & \kappa_{12} & \kappa_{13} \\ 0 & \kappa_{22} & \kappa_{23} \\ 0 & 0 & \kappa_{33} \end{bmatrix} \tag{85}$$

In the paper [50], instead of the full $\omega$ matrix, the $\omega'$ matrix containing only the two initial terms of the series (40) was used

$$\omega' = I_3 + \kappa_{\mathrm{U}} = \begin{bmatrix} 1+\kappa_{11} & \kappa_{12} & \kappa_{13} \\ 0 & 1+\kappa_{22} & \kappa_{23} \\ 0 & 0 & 1+\kappa_{33} \end{bmatrix} \tag{86}$$

In the paper [51], the $\omega$ matrix is approximated by an $\omega''$ matrix of the form

$$\begin{aligned} \omega'' &= I_3 + \kappa_{\mathrm{U}} + \kappa_{\mathrm{U}}^2 + \kappa_{\mathrm{D}}^3 + \kappa_{\mathrm{D}}^4 + \kappa_{\mathrm{D}}^5 + \ldots \\ &= \begin{bmatrix} \dfrac{1}{1-\kappa_{11}} & \kappa_{12}(1+\kappa_{11}+\kappa_{22}) & \begin{matrix}\kappa_{13}(1+\kappa_{11}+\kappa_{33}) \\ +\kappa_{12}\kappa_{23}\end{matrix} \\ 0 & \dfrac{1}{1-\kappa_{22}} & \kappa_{23}(1+\kappa_{22}+\kappa_{33}) \\ 0 & 0 & \dfrac{1}{1-\kappa_{33}} \end{bmatrix} \end{aligned} \tag{87}$$

where $\kappa_{\mathrm{D}} = \mathrm{diag}(\kappa_{11},\kappa_{22},\kappa_{33})$. The approximation of the $\omega$ matrix using the $\omega''$ matrix is better than using the $\omega'$ matrix, but it is still worse than using the full $\omega$ matrix. This is because the $\kappa_{\mathrm{D}}$ matrix is used instead of the $\kappa_{\mathrm{U}}$ matrix in the higher expressions of expansion (87).

## Appendix C: Relation between $I_{ss}$ and $I_\delta(t)$

Excitation of fluorescence with continuous light of photon flux density $I_{\mathrm{ex}}$ is equivalent to excitation with a compact sequence of rectangular pulses, each of small width $\Delta t$ and photon density $J_{\mathrm{ex}}$. The values of $J_{\mathrm{ex}}$ and $I_{\mathrm{ex}}$ are related by the expression

$$J_{\mathrm{ex}} = I_{\mathrm{ex}}\,\Delta t \tag{88}$$

The fluorescence intensity $I_{\mathrm{ss}}$ observed with continuous excitation is as if all fluorescence quanta generated by each individual excitation pulse were emitted within a single $\Delta t$ time segment

$$I_{ss} = \frac{1}{\Delta t}\int_0^\infty I_\delta(t)\,dt \tag{89}$$

Finally, after taking into account (88), we can write

$$\int_0^\infty I_\delta(t)\,dt = \frac{J_{\mathrm{ex}}}{I_{\mathrm{ex}}} I_{ss} \tag{90}$$

## Appendix D: Consistency of expression (68) with an earlier expression obtained in the paper [45]

If the solution contains only one component, the matrix $\Phi(t)$ reduces to a single element $\Phi_{11}(t)$. Defined in [45], the function $S^{(1)}(t)$ has the meaning of an excitation survival function among the originally excited molecules of component 1, which means that

$$S^{(1)}(t) = \frac{\Phi_{11}(t)}{\Phi_{11}(0)} \tag{91}$$

Hence, we have

$$\hat{\Phi}_{11}(s) = \Phi_{11}(0)\,\hat{S}^{(1)}(s) \tag{92}$$

Given that from equation (43) follows $\phi_{11} = \hat{\Phi}_{11}(s=0)$, we can write

$$\phi_{11} = \Phi_{11}(0)\,\hat{S}^{(1)}(s=0) \tag{93}$$

Thus, there is a relation

$$\hat{\Phi}_{11}(s) = \phi_{11}\,\frac{\hat{S}^{(1)}(s)}{\hat{S}^{(1)}(s=0)} \tag{94}$$

After inserting (92) and (94) into (68) we get

$$\hat{I}_\delta(s) = g\,J_{\mathrm{ex}}\,C\,\phi_1^\star\,F_1\,\frac{\Phi_{11}(0)\,\hat{S}^{(1)}(s)}{1-\kappa_{11}\dfrac{\hat{S}^{(1)}(s)}{\hat{S}^{(1)}(s=0)}} \tag{95}$$

If we consider only the primary fluorescence of this system, then based on equation (48) we can write

$$I_\delta^{(1)}(t) = g\,J_{\mathrm{ex}}\,C\,\phi_1^\star\,F_1\,\Phi_{11}(t) \tag{96}$$

Hence, we see that

$$I_{\delta 0} = g\,J_{\mathrm{ex}}\,C\,\phi_1^\star\,F_1\,\Phi_{11}(t=0) \tag{97}$$

is the value of $I_\delta^{(1)}(t)$ at $t=0$. The same value of $I_{\delta 0}$ is also the initial value in expression (95), since taking into account secondary and higher order emissions does not affect the fluorescence intensity at $t=0$. This allows equation (95) and thus equation (68) to be written in the form of

$$\hat{I}_\delta(s) = I_{\delta 0}\,\frac{\hat{S}^{(1)}(s)}{1-\kappa_{11}\,\hat{S}^{(1)}(s)\big/\hat{S}^{(1)}(s=0)} \tag{98}$$

which is consistent with equation (30) in the paper [45].

# Supplementary materials to the work: “Multistep reversible excitation transfer in a multicomponent rigid solution: I. Calculation of steady-state and time-resolved fluorescence intensities”

**Józef Kuśba**[1]
*Independent researcher. Retired at Faculty of Applied Physics and Mathematics, Gdańsk University of Technology, Gdańsk, Poland*

## 1 FORTRAN code for calculation the function *M(α,β,γ,m)* given by Eq. (35)

Function $M(\alpha,\beta,\gamma,m)$ is defined as

$$M(\alpha,\beta,\gamma,m)=\frac{\alpha+\beta}{1-e^{-(\alpha+\beta)}}\frac{\gamma}{2}\int_0^1 e^{-\beta u_0}\int_0^1 e^{-\alpha u}\left[\mathrm{Ei}\left(-\gamma\sqrt{m^2+(u-u_0)^2}\right)-\mathrm{Ei}\left(-\gamma\,|\,u-u_0\,|\right)\right]du\,du_0 \tag{1}$$

It seems that the simplest way to calculate the values of this function is to calculate the double integral numerically. In the case of $m>>1$, you can use the faster method given below. If $m>>1$ then one obtains

$$M(\alpha,\beta,\gamma,m)=\frac{\alpha+\beta}{1-e^{-(\alpha+\beta)}}\frac{\gamma}{2}\int_0^1 e^{-\beta u_0}\int_0^1 e^{-\alpha u}\left[\mathrm{Ei}\left(-m\gamma\right)-\mathrm{Ei}\left(-\gamma\,|\,u-u_0\,|\right)\right]du\,du_0 \tag{2}$$

After analytical transformations, equation (2) can be written in the form [1,2]

$$\begin{aligned}M(\alpha,\beta,\gamma)&=\frac{\alpha+\beta}{2\alpha\beta}\frac{\left(1-e^{-\alpha}\right)\left(1-e^{-\beta}\right)}{1-e^{-(\alpha+\beta)}}\left[\gamma\,\mathrm{Ei}(-m\gamma)-\gamma\,\mathrm{Ei}(-\gamma)\right]\\&+\frac{1}{2\left(1-e^{-(\alpha+\beta)}\right)}\left[\chi(\alpha,\gamma)+\chi(\beta,\gamma)+e^{-\beta}\psi(\alpha,\gamma)+e^{-\alpha}\psi(\beta,\gamma)\right]\end{aligned} \tag{3}$$

where

$$\chi(x,\gamma)=\frac{\gamma}{x}\left[G(-\gamma)-G(-(\gamma+x))\right] \tag{4}$$

$$\psi(x,\gamma)=\frac{\gamma e^{-x}}{x}\left[G(-\gamma)-G(-(\gamma-x))\right] \tag{5}$$

$$G(x)=\mathrm{Ei}(x)-\ln|\,x\,| \tag{6}$$

### 1.1 Function MP(alpha,beta,gamma,m) in the case *m* >> 1

```
real(8) function MP(alpha,beta,gamma,m)
    ! calculates values of the function M according to the formulae given in
    ! A. Budo and I. Ketskemety, Acta Phys. Hung. 14 (1962) 167-176.
    use exprl_int
```

[1] ORCID: 0000-0002-9697-0751. Electronic mail: jozkusba@pg.edu.pl

```
    use ei_int
    implicit none
    real(8),intent(in)      :: alpha    ! Napierian absorbance of the sample for the excitation
                                        ! wavelength
    real(8),intent(in)      :: beta     ! Napierian absorbance of the sample for the observation
                                        ! wavelength
    real(8),intent(in)      :: gamma    ! Napierian absorbance of the sample for given wavelength
                                        ! from the spectra overlapping area
    real(8),intent(in)      :: m        ! R/l
    real(8)                 :: w1,w2,ea,eb,eab1
    real(8),external        :: chi,psi   !
    if (gamma.eq.0d0) then
        MP=0d0
        return
    endif
    ea=dexp(-alpha)
    eb=dexp(-beta)
    eab1=1d0-ea*eb
    ! exprl(x)=(exp(x)-1)/x     function from IMSL library
    ! Ei    function from IMSL library
    ! chi   function from file chi.f90
    ! psi   function from file psi.f90
    w1=exprl(-alpha)*exprl(-beta)*gamma/exprl(-alpha-beta)*(Ei(-m*gamma)-Ei(-gamma))
    w2=(chi(alpha,gamma)+chi(beta,gamma)+eb*psi(alpha,gamma)+ea*psi(beta,gamma))/eab1
    MP=(w1+w2)/2d0
    end function MP
```

### 1.2 Function chi(x,y)

```
real(8) function chi(x,y)
    use ei_int
    implicit none
    real(8),intent(in)      :: x,y
    if (y.eq.0d0) then
        chi=0d0
    elseif (x.eq.0d0) then
        chi=1.0d0-dexp(-y)
    else
        chi=y/x*(Ei(-y)-Ei(-x-y)-dlog(y)+dlog(x+y))
    endif
    return
    end function chi
```

### 1.3 Function psi(x,y)

```
real(8) function psi(x,y)
    use ei_int
    implicit none
    real(8),intent(in)      :: x,y
    integer                 :: n
    real(8)                 :: w1,w2,s
    if (y.eq.0d0) then
        psi=0d0
    elseif (x.eq.0d0) then
        psi=dexp(-y)-1d0
    elseif (x.eq.y) then
```

```
        if (x.gt.4d1) then
            psi=0d0
        else
            n=1
            s=1d0
            w1=0d0
            w2=1d0
            do while (dabs(w1-w2).ge.1.0d-10)
                w2=w1
                s=-s*y/n
                w1=w1+s/n
                n=n+1
            enddo
            psi=w1*dexp(-y)
        endif
    else
        psi=y/x*dexp(-x)*(Ei(-y)-Ei(x-y)+dlog(dabs(x-y))-dlog(y))
    endif
    return
end function psi
```

### 1.4 Function Ei(x)

```
real(8) function Ei(x)
! This code can be used if one has no access to IMSL library
! x has to be negative here
    implicit none
    real(8),intent(in)      :: x
    integer                 :: n
    real(8),parameter       :: C        = 0.577215664901532860606d0     ! Euler's constant
    real(8)                 :: w1,w2,s
!    if (x.ge.0d0) then
!        write(6,*) 'Nonnegative argument of the function Ei'
!        stop
!    else
    if (x.lt.-10d0) then
        Ei=dexp(x)/x*(1d0+1d0/x*(1d0+2d0/x*(1d0+3d0/x*(1d0+4d0/x*(1d0+5d0/x)))))
        return
    else
        n=2
        w1=x
        s=x*x/n
        w2=w1+s/n
        do while (dabs(w1-w2).ge.1d-10)
            n=n+1
            w1=w2
            s=s*x/n
            w2=w1+s/n
        enddo
        Ei=C+dlog(dabs(x))+w2
    endif
    return
end function Ei
```

# 2 MATHCAD code for the function *M(α,β,γ,m)* given by Eq. (35) in the case *m* >> 1

## 2.1 Function MP(alpha,beta,gamma,m)

```
MP(α, β, γ, m) := | if γ = 0
                  |    | MP ← 0
                  |    | return MP
                  | otherwise
                  |    | ea ← exp(−α)
                  |    | eb ← exp(−β)
                  |    | eab1 ← 1 − ea·eb
                  |    | w1 ← exprl(−α)·exprl(−β)·[γ / exprl(−α − β)]·(Ei(−m·γ) − Ei(−γ))
                  |    | w2 ← (chi(α,γ) + chi(β,γ) + eb·psi(α,γ) + ea·psi(β,γ)) / eab1
                  |    | MP ← (w1 + w2) / 2
                  | MP
```

## 2.2 Function exprl(x)

```
exprl(x) := | exprl ← (exp(x) − 1) / x  if |x| > 0.1
            | otherwise
            |    | n ← 1
            |    | s ← 1
            |    | w ← 1
            |    | while |s| > 10^−20
            |    |    | n ← n + 1
            |    |    | s ← s·(x / n)
            |    |    | w ← w + s
            |    | exprl ← w
            | exprl
```

## 2.3 Function chi(x,y)

```
chi(x,y) := | if y = 0
            |    | chi ← 0
            |    | return chi
            | if x = 0
            |    | chi ← 1 − exp(−y)
            |    | return chi
            | chi ← (y/x)·(Ei(−y) − Ei(−x − y) − ln(y) + ln(x + y))  otherwise
            | chi
```

## 2.4 Function psi(x,y)

```
psi(x,y) := | if y = 0
            |    | psi ← 0
            |    | return psi
            | if x = 0
            |    | psi ← exp(−y) − 1
            |    | return psi
            | if x = y
            |    | psi ← 0  if x > 40
            |    | otherwise
            |    |    | n ← 1
            |    |    | s ← 1
            |    |    | w1 ← 0
            |    |    | w2 ← 1
            |    |    | while |w1 − w2| ≥ 10^−10
            |    |    |    | w2 ← w1
            |    |    |    | s ← −s·(y/n)
            |    |    |    | w1 ← w1 + s/n
            |    |    |    | n ← n + 1
            |    |    | psi ← w1·exp(−y)
            |    | return psi
            | psi ← (y/x)·exp(−x)·(Ei(−y) − Ei(x − y) + ln(|x − y|) − ln(y))  otherwise
            | psi
```

### 2.5 Function Ei(x)

In our calculations we used here $xg = -10$

$$
Ei(x) := \left|\begin{array}{l}
Ei \leftarrow \frac{\exp(x)}{x}\cdot\left[1+\frac{1}{x}\cdot\left[1+\frac{2}{x}\cdot\left[1+\frac{3}{x}\cdot\left[1+\frac{4}{x}\cdot\left[1+\frac{5}{x}\cdot\left(1+\frac{6}{x}\right)\right]\right]\right]\right]\right] \quad \text{if } x < xg \\
\text{otherwise} \\
\quad \left|\begin{array}{l}
n \leftarrow 2 \\
w1 \leftarrow x \\
s \leftarrow \frac{x\cdot x}{n} \\
w2 \leftarrow w1 + \frac{1}{n}\cdot s \\
\text{while } \left(|w1 - w2| \geq 10^{-10}\right) \\
\quad \left|\begin{array}{l}
n \leftarrow n + 1 \\
w1 \leftarrow w2 \\
s \leftarrow s\cdot\frac{x}{n} \\
w2 \leftarrow w1 + \frac{s}{n}
\end{array}\right. \\
Ei \leftarrow 0.57721566490153286060 6 + \ln(|x|) + w2
\end{array}\right. \\
Ei
\end{array}\right.
$$